\documentclass[twocolumn,aps,tightenlines,floatfix,showpacs]{revtex4}
\usepackage{amsmath}
\usepackage[dvips]{graphicx}
\usepackage{hhline}
\usepackage{epsfig}

\newcommand{\beq}{\begin{equation}}
\newcommand{\eeq}{\end{equation}}
\newcommand{\beqn}{\begin{eqnarray}}
\newcommand{\eeqn}{\end{eqnarray}}
\newcommand{\bsig}{\mbox{\boldmath{$\sigma$}}}

\begin{document}

\title{Neutrinoless double beta decay of ${}^{48}$Ca in the shell model: Closure versus nonclosure approximation}

\author{R.A.~Sen'kov and M.~Horoi}

\affiliation{Department of Physics, Central Michigan University, Mount Pleasant, 
Michigan 48859, USA}


\begin{abstract}
Neutrinoless double-$\beta$ decay ($0\nu\beta\beta$) is a unique process 
that could reveal physics beyond the Standard Model. 
Essential ingredients in the analysis of $0\nu\beta\beta$ 
rates are the associated nuclear matrix elements. Most of the 
approaches used to calculate these matrix elements rely on the closure approximation. 
Here we analyze the light neutrino-exchange matrix elements of ${}^{48}$Ca 
$0\nu\beta\beta$ decay and test the closure approximation in a shell-model
approach. We calculate the $0\nu\beta\beta$ nuclear matrix elements for 
${}^{48}$Ca using both the closure approximation and a nonclosure approach, 
and we estimate the uncertainties associated with the closure approximation. 
We demonstrate that the nonclosure approach has excellent convergence
properties which allow us to avoid unmanageable computational cost. 
Combining the nonclosure and closure approaches we propose a 
new method of calculation for $0\nu\beta\beta$ decay rates which can be 
applied to the $0\nu\beta\beta$ decay rates of heavy nuclei, such as
${}^{76}$Ge or ${}^{82}$Se. 
\end{abstract}

\maketitle

\section{Introduction}

Neutrinoless double-$\beta$ decay ($0\nu\beta\beta$), 
if observed, would prove that neutrinos 
are Majorana fermions, an important milestone in the search for physics 
beyond the Standard Model \cite{sv82}. In addition, 
one could extract more information about the nature of the decay 
mechanism and possibly determine the light neutrino mass hierarchy and 
the lightest neutrino mass \cite{ves12,tomoda}, provided that the 
associated nuclear matrix elements (NME) are calculated with good 
accuracy \cite{ves12,prc13,prl13,prc10,prc07}.

There are many possible mechanisms that could contribute to the
$0\nu\beta\beta$ decay process \cite{ves12,prc13}, and some of the 
associated matrix elements were investigated by using several approaches, 
including the quasiparticle random phase approximation (QRPA) \cite{ves12}, 
the interacting shell model \cite{prl100,prc13}, the interacting boson model \cite{iba-2}, the generator coordinate method \cite{gcm}, and the projected Hartree-Fock Bogolibov
model \cite{phfb}. 
With the exception of the QRPA, all other methods entail  
 using the closure approximation 
\cite{prc10}. Some older \cite{pv90,muto94} and more recent 
\cite{simvo11} analyses suggest that the deviation of the NME for the 
light neutrino-exchange mechanism from the closure approximation result 
should be small, but a full analysis of this deviation within the shell 
model is not yet available. In addition, the QRPA analysis is affected 
by uncertainties due to the $g_{pp}$ factor used to tune the residual
interaction. For example, results from Ref. \cite{muto94} indicate 
a deviation of 
about up to 10\% between closure and nonclosure NME, but its magnitude 
and sign depend on the choice of $g_{pp}$. The only shell-model analysis 
going beyond the closure approximation that we are aware of was done in 
Ref. \cite{pv90} for 
${}^{48}$Ca using a model space consisting of only the $f_{7/2}$ orbital. 
This model space is known to be insufficient for a good description of 
the NME due to the missing spin-orbit partner orbital $f_{5/2}$, which
significantly reduces the 
Gamow-Teller strength. The authors of Ref. \cite{pv90} report very small changes of 
the NME from closure to nonclosure, and in most cases the magnitude of 
the nonclosure results is slightly smaller than the magnitude of the 
closure result.

In this paper we analyze and compare the closure and nonclosure NME 
for the $0\nu\beta\beta$ decay of ${}^{48}$Ca using a shell-model 
approach in the full $pf$ shell \cite{prc10,prc13}. For the analysis we 
used the GXPF1A interaction \cite{gxpf1,gxpf1a}. This analysis requires 
knowledge of a large number of one-body transition densities 
connecting the ground states of the initial and final states of ${}^{48}$Ca and 
${}^{48}$Ti, respectively, with  states of the intermediate 
nucleus ${}^{48}$Sc. 
The total number of states in ${}^{48}$Sc with angular momentum smaller than $J=7$  is about 100000. This is still an unmanageable task. 
However, we show that using only a few hundred states of each $J$ suffices 
to get accurate NME. In order to validate our results we also analyzed 
the $0\nu\beta\beta$ NME of the ``fictitious" decays of ${}^{44}$Ca 
and ${}^{46}$Ca, for which a full account of all relevant states in 
the intermediate nucleus ${}^{48}$Sc is possible. We find that the 
nonclosure NME always increases relative to its closure value by 
about 10\%.

The paper is organized as follows. Section II gives a brief description 
of the light neutrino exchange $0\nu\beta\beta$ NME relevant for the 
distinction between the nonclosure approach and the closure 
approximation. Section III provides a brief description of the 
closure approximation. Section IV describes the approach we use to 
obtain the nonclosure results and outlines new mixed methods that use 
the closure approach to accelerate the convergence. In Sect. V we analyze 
the numerical results, and Sec. VI is devoted to conclusions and 
outlook. Details of the calculations are shown in the appendices. 

\section{The nuclear matrix element}

The decay rate for a $0\nu\beta\beta$ decay process, under the assumption  
that the light neutrino-exchange mechanism dominates \cite{ves12,prc13}, 
can be written as
\beq
\left[ T^{0\nu}_{1/2} \right]^{-1} = G^{0\nu} | M^{0\nu} |^2 
\left(\frac{\langle m_{\beta \beta}\rangle}{m_e}\right )^2.
\eeq
Here $G^{0\nu}$ is the phase-space factor \cite{kipf12}, 
$M^{0\nu}$ is the nuclear matrix element, and the effective neutrino 
mass $ \langle m_{\beta \beta}\rangle$ is defined by the neutrino 
mass eigenvalues $m_k$ and the elements of neutrino mixing matrix 
$U_{ek}$ \cite{ves12},
\beq
\langle m_{\beta \beta}\rangle = \left| \sum_k m_k U^2_{ek} \right|.
\eeq
The nuclear matrix element  $M^{0\nu}$ is usually presented as a sum 
of Gamow-Teller (GT), Fermi (F), and Tensor (T) \cite{corr1} nuclear matrix 
elements (see, for example, Ref. \cite{prc10}), 
\beq \label{nme1}
M^{0\nu} = M^{0\nu}_{GT} - \left( \frac{g_{V}}{g_{A}} \right)^2  
M^{0\nu}_{F} + M^{0\nu}_{T},
\eeq
where $g_{V}$ and $g_{A}$ are the vector and axial constants 
correspondingly; in our calculations we use  $g_{V}=1$ and $g_{A}=1.254$.

The nuclear matrix elements in Eq. (\ref{nme1}) describe the transition from 
an initial nucleus $|i\rangle=|0^+_i\rangle$ to a final nucleus 
$|f\rangle=|0^+_f\rangle$, and they can be presented as a sum over 
intermediate nuclear states $| \kappa \rangle=|J^\pi_\kappa \rangle$ 
with certain angular momentum $J_\kappa$, parity $\pi$, and 
energy $E_\kappa$
\beq \label{nme2}
M^{0\nu}_{\alpha}=\sum_{\kappa} \sum_{1234} \langle 1 3 | {\cal O}_{\alpha} | 2 4\rangle
\langle f |  \hat{c}^\dagger_{3} {\hat{c}}_4 | \kappa \rangle
\langle \kappa |  \hat{c}^\dagger_{1} {\hat{c}}_2 | i \rangle,
\eeq
where operators ${\cal O}_{\alpha}$, $\alpha=\{ GT, F, T \}$, contain 
neutrino potentials, spin and isospin operators, and the explicit dependence 
on the intermediate state energy $E_\kappa$. They are given by
\beq
\begin{aligned} \label{op}
{\cal O}_{GT} = & \tau_{1-} \tau_{2-} \; (\bsig_1 \cdot \bsig_2) \; H_{GT}(r, E_\kappa), \\
{\cal O}_{F} = & \tau_{1-} \tau_{2-} \; H_{F}(r, E_\kappa), \\
{\cal O}_{T} = & \tau_{1-} \tau_{2-} \; S_{12}\; H_{T}(r,  E_\kappa),
\end{aligned}
\eeq
with $S_{12}=3(\bsig_1 \cdot {\bf n})(\bsig_2 \cdot {\bf n})- (\bsig_1 \cdot \bsig_2)$, 
${\bf r}={\bf r}_1-{\bf r}_2$, $r= | {\bf r} |$, and ${\bf n}={\bf r}/r$. 
The neutrino potentials, $H_{\alpha}(r, E_\kappa)$, are integrals over 
the neutrino exchange momentum, $q$,
\beq \label{pot} 
H_{\alpha}(r, E_\kappa)=\frac{2 R}{\pi} \int_0^\infty
\frac{ f_{\alpha}(q r) h_{\alpha}(q^2) q d q}{q+E_\kappa - (E_i + E_f)/2},
\eeq
where $f_{GT, F}(q r)=j_0(q r)$ and $f_{T}(q r)=j_2(q r)$ are spherical 
Bessel functions. The nuclear radius $R=1.2\times A^{1/3}\,{\rm fm}$ 
was introduced 
to make the neutrino potentials dimensionless (and since the phase-space factor
$G^{0\nu}$ contains $1/R^2$ the final transition probability 
does not depend on $R$). The form factors $h_{\alpha}(q^2)$ are defined 
in Appendix \ref{ap1} and they include vector and axial nucleon form factors 
that take into account nucleon size effects. 
Calculation details for two-body matrix elements, 
$\langle 1 3 | {\cal O}_{\alpha} | 2 4\rangle$, are discussed in Appendix
\ref{ap4}. 
Let us note that the two-body wave functions 
in the matrix elements (\ref{nme2}) are not antisymmetrized, as one would expect 
for nuclear two-body matrix elements. They should be understood as
\beq 
|2 4\rangle = |2\rangle \cdot |4\rangle\; \mbox{ and } \; 
|1 3\rangle = |1\rangle \cdot |3\rangle,
\eeq 
where 1, 2, 3, and 4 represent single-nucleon quantum numbers  
(for example, $1=\{ \tau_{1 z},n_1,l_1,j_1,\mu_1 \}$ and so on).


Appendices \ref{ap2}, \ref{ap3}, and \ref{ap4} provide expressions
for the nuclear matrix elements (\ref{nme2}) by considering rotational
symmetry and isospin invariance.

\section{The closure approximation}

If one replaces the energies of the intermediate states in 
Eq. (\ref{pot}) by an average constant value one gets the closure 
approximation,
\beq \label{eq8}
\left[ E_\kappa - (E_i + E_f)/2 \right] \rightarrow \langle E \rangle.
\eeq
The operators 
${\cal O}_\alpha \rightarrow \tilde{{\cal O}}_\alpha \equiv {\cal O}_\alpha(\langle E \rangle)$ 
become energy independent and the sum over the intermediate states in 
the nuclear matrix element (\ref{nme2}) can be taken explicitly by 
using the completeness relation
\beq \label{tbtd}
\sum_\kappa 
\langle f |  \hat{c}^\dagger_{3} {\hat{c}}_4 | \kappa \rangle
\langle \kappa |  \hat{c}^\dagger_{1} {\hat{c}}_2 | i \rangle =
\langle f |  \hat{c}^\dagger_{3} {\hat{c}}_4  \hat{c}^\dagger_{1} {\hat{c}}_2 | i \rangle.
\eeq
The advantage of this approximation is significant, because it eliminates the 
need of calculating a very large number of states in the intermediate 
nucleus, which could be computationally challenging, especially for heavy
systems. One needs only to calculate the two-body transition densities (\ref{tbtd}) between the initial and the final nuclear states. 
This approximation is very good because the values of $q$ 
that dominate the matrix elements are of the order of $100-200$ MeV, while 
the relevant excitation energies are only of the order of 10 MeV.
The obvious difficulty related to this approach is that we have to find 
a reasonable value for this average energy, $\langle E \rangle$, which 
can effectively represent the contribution of all the intermediate states.
This average energy needs to account also for the symmetric part
of the two-body matrix elements, 
$\langle 1 3 | {\cal O}_{\alpha} | 2 4\rangle$, in Eq. (\ref{nme2}).
Indeed, the two-body wave functions $|1 3\rangle$ and $|2 4\rangle$ are 
not antisymmetric; by replacing the energies of the intermediate states 
with a constant, only the antisymmetric part of these matrix elements 
is taken into account.

The uncertainty in the value of the nuclear matrix elements is 
related to our inability to derive the average energy, 
$\langle E \rangle$,  associated
with the closure approximation. Fortunately, the nuclear matrix 
elements are not very sensitive to the value of this average energy 
(with the uncertainty being estimated to be about 10\%; see, for example, 
\cite{prc10}). Such weak dependence on the average energy originates from 
the large value of typical momentum of the virtual neutrino 
[see Eq. (\ref{pot})],
which is $\sim 1\,{\rm fm}^{-1}$ ($\sim 200\,{\rm MeV}$), i.e., 
much larger than the typical nuclear excitations. 

\section{Nonclosure and mixed methods}

In the nonclosure approach one needs to calculate the 
sum in Eq. (\ref{nme2}) explicitly, which is an obvious 
challenge due to the large number of intermediate states 
$|\kappa \rangle$. For the case of ${}^{48}$Ca in the $fp$ model space 
there are about $10^5$ intermediate states; it is
 extremely difficult to find and include all these states.

Let us introduce a cutoff energy $E$ to investigate the convergence 
of the sum over $\kappa$ in Eq. (\ref{nme2})  (where here and below 
the sum over repeated indices $\{1,2,3,4\}$ is omitted): 
\beq\label{nme3}
M^{0\nu}_{\alpha}(E)=\sum_{E_\kappa < E} 
\langle 1 3 | {\cal O}_{\alpha} | 2 4\rangle
\langle f |  \hat{c}^\dagger_{3} {\hat{c}}_4 | \kappa \rangle
\langle \kappa |  \hat{c}^\dagger_{1} {\hat{c}}_2 | i \rangle.
\eeq
Alternatively, we can use a cutoff on the number of states, $N$, 
calculating the sum only for $\kappa < N$. 
At the limit of large cutoff energies $M^{0\nu}_{\alpha}(E)$
approaches the exact value of the nuclear matrix element (\ref{nme2}).

The difference between the closure and
nonclosure calculations originates mainly from the low-lying excitation
energies. The intermediate and higher energies cannot produce much of a
difference, because with increase of the excitation energy the  
one-body matrix elements rapidly become very small.
Based on this observation, we will use the nonclosure approach for low energies, which we can manage within the framework of the 
standard shell model. For the higher excitation energies, we will use 
the closure approximation, which is also manageable. 
To proceed further we introduce the sum similar to Eq. (\ref{nme3}) for the closure approximation:
\beq\label{nme4}
{\cal M}^{0\nu}_{\alpha}(E)=\sum_{E_\kappa < E} 
\langle 1 3 | \tilde{{\cal O}}_{\alpha} | 2 4\rangle
\langle f |  \hat{c}^\dagger_{3} {\hat{c}}_4 | \kappa \rangle
\langle \kappa |  \hat{c}^\dagger_{1} {\hat{c}}_2 | i \rangle.
\eeq
The difference between Eqs. (\ref{nme3}) and (\ref{nme4}) is that for
the nonclosure approach the operators ${\cal O}_\alpha$ in Eq. (\ref{op}) 
are functions
of the excitation energy $E_\kappa$, while for the closure approximation 
the same operators $\tilde{{\cal O}}_\alpha$ are functions
of the average energy $\langle E \rangle$ [see the energy substitution
given by Eq. (\ref{eq8})]. At large cutoff energies,
$E\rightarrow \infty$,
\beq \label{eq12}
{\cal M}^{0\nu}_{\alpha}(E) \rightarrow {\cal M}^{0\nu}_{\alpha}(\infty)=
\langle 1 3 | \tilde{{\cal O}}_{\alpha} | 2 4\rangle
\langle f |  \hat{c}^\dagger_{3} {\hat{c}}_4 \hat{c}^\dagger_{1} {\hat{c}}_2 
| i \rangle,
\eeq
we get an ``exact value" in the framework of the closure approximation.

To avoid disadvantages of both approaches we 
propose an interpolation method which combines both the nonclosure
and closure approaches, by introducing the mixed NME
\beq\label{nme5}
{\bar M^{0\nu}_{\alpha}}(E)=
{M}^{0\nu}_{\alpha}(E)-{\cal M}^{0\nu}_{\alpha}(E)+{\cal M}^{0\nu}_{\alpha}(\infty).
\eeq 
We expect that this mixed NME, ${\bar M^{0\nu}_{\alpha}}(E)$, will converge much faster with the cutoff energy than the nonclosure, $M^{0\nu}_{\alpha}(E)$, and closure, ${\cal M}^{0\nu}_{\alpha}(E)$, matrix elements separately. At
higher excitation energies these two NME will behave similarly, and the energy
dependence will cancel out. We also expect that the mixed NME, 
Eq. (\ref{nme5}), will have  much weaker dependence on the average 
energy $\langle E \rangle$ than the
pure closure NME; at least this dependence should weaken when the cutoff energy increases. 
It should be also mentioned that calculating ${\cal M}^{0\nu}_{\alpha}(E)$ and 
${\bar M^{0\nu}_{\alpha}}(E)$ does not require more computational effort than calculating 
the energy-dependent nonclosure NME, $M^{0\nu}_{\alpha}(E)$, for a given energy cutoff. 
${\cal M}^{0\nu}_{\alpha}(\infty)$ can be calculated by using 
Eq. (\ref{eq12}) (the details of which are described in Ref. \cite{prc10}).
   
\section{Results}

\begin{figure}
\begin{center}
\includegraphics[width=0.48\textwidth]{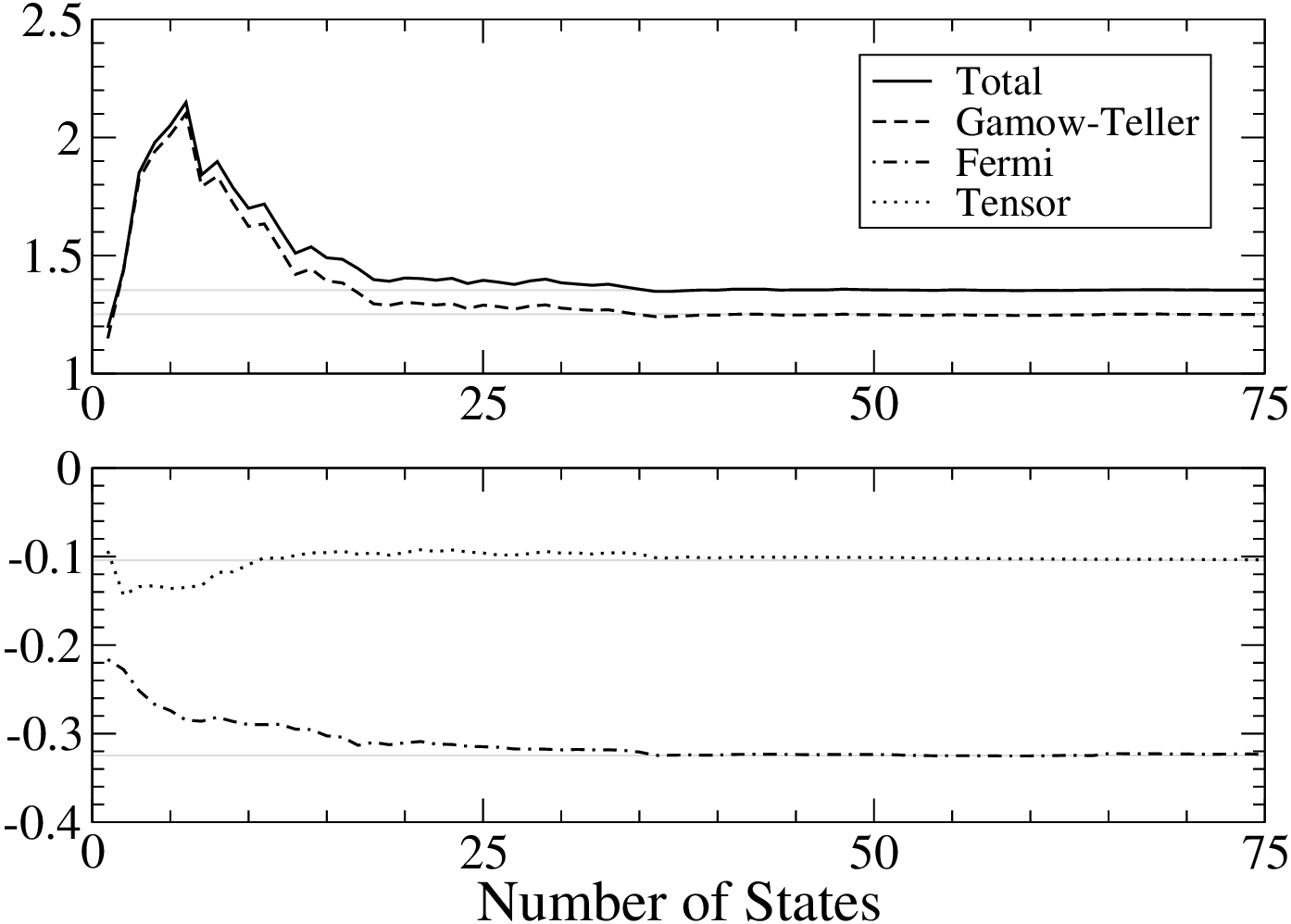}
\caption{Convergence of closure NME ${\cal M}^{0\nu}_{\alpha}(E)$ 
with the number of intermediate states $|\kappa \rangle$ for the
fictitious $0\nu\beta\beta$ decay of ${}^{44}$Ca: 
Total (solid curve) and Gamow-Teller (dashed curve) (upper panel) and
Fermi (dash-dotted curve) and Tensor (dotted curve) (lower panel).\\  
}\label{fig1}
\end{center}
\end{figure}

\begin{figure}
\begin{center}
\includegraphics[width=0.48\textwidth]{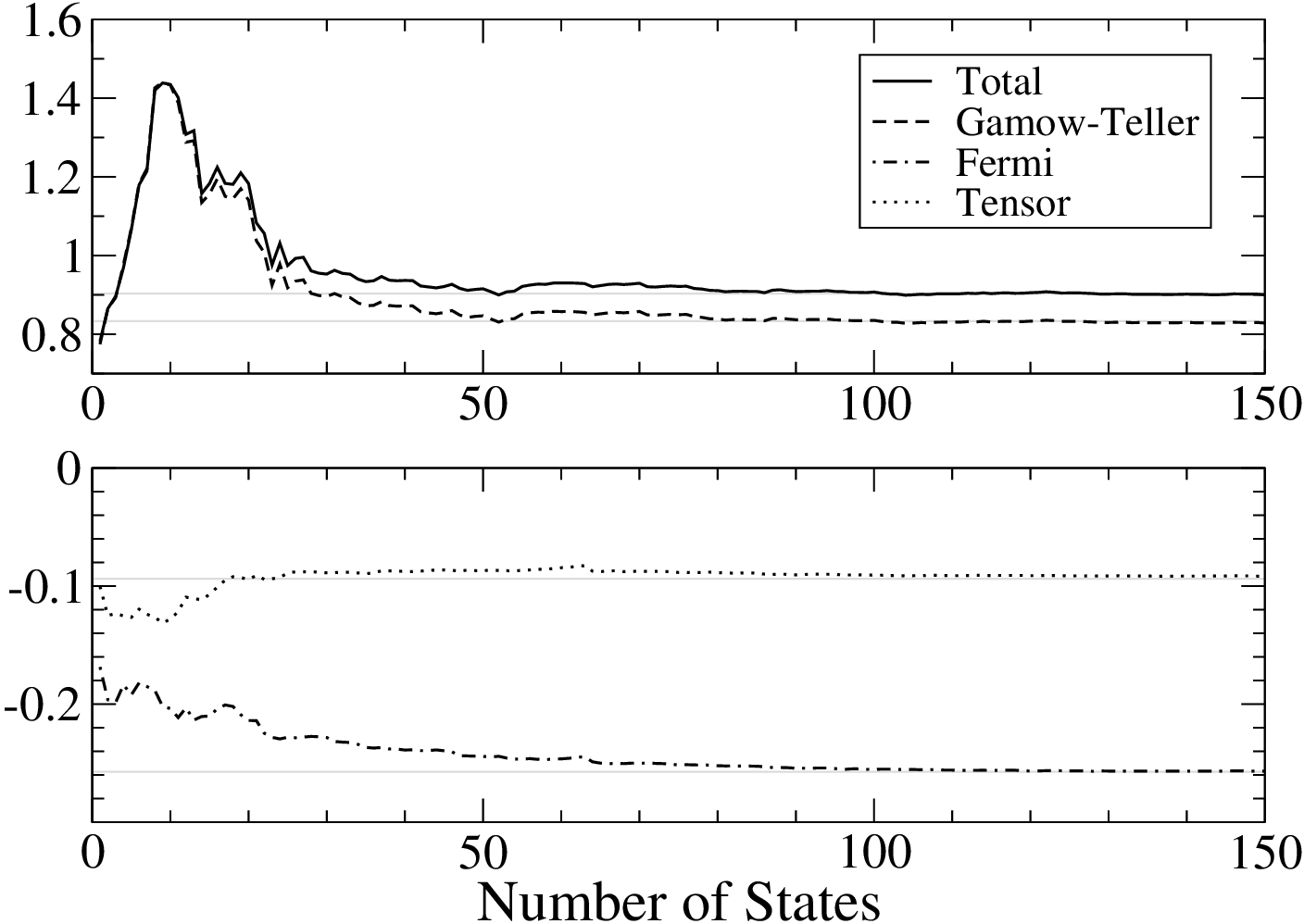}
\caption{ 
Same as Fig. \ref{fig1} for the
fictitious $0\nu\beta\beta$ decay of ${}^{46}$Ca.\\}\label{fig2}
\end{center}
\end{figure}

Figures \ref{fig1} and \ref{fig2} present the closure NME
${\cal M}^{0\nu}_{\alpha}(E)$ for the 
fictitious $0\nu\beta\beta$ decay cases of ${}^{44}$Ca and ${}^{46}$Ca. 
We calculated NME for these two cases only to demonstrate
the convergence of the corresponding nuclear matrix elements 
with the increase of the cutoff energy. 
We could check our code by comparing
with the NME calculated with a totally different method \cite{prc10,prc13}.  
The one-body transition densities 
($\langle f |  \hat{c}^\dagger_{3} {\hat{c}}_4 | \kappa \rangle$ and 
$\langle \kappa |  \hat{c}^\dagger_{1} {\hat{c}}_2 | i \rangle$)
 were calculated with the NUSHELLX code \cite{nushellx}, and we developed our 
code for the two-body matrix elements. 
We used the 
GXPF1A two-body interaction \cite{gxpf1,gxpf1a} in the $pf$ model space.
In the calculations we used $\langle E \rangle=7.72$ MeV, and 
we also included the short-range correlations (SRC) parametrization based
on the AV18 potential and the standard nucleon finite-size effects \cite{prc10}. 
The horizontal lines
represent the ``exact values", ${\cal M}^{0\nu}_{\alpha}(\infty)$. 
One can see how the NME converge to their
exact values: for ${}^{46}$Ca it is enough to take into account  
about 50 states (instead of $\sim 20\,000$) and for ${}^{44}$Ca 
about 25 states are needed to obtain an accuracy better 
than 1\% for the total NME.
We should also mention that for ${}^{44}$Ca and ${}^{46}$Ca 
we were able to include all the states in the intermediate nucleus, and
we got the same results as using the traditional nonclosure approach \cite{prc10,prc13} [see, e.g., Eq. (\ref{tbtd})].

\begin{figure}
\begin{center}
\includegraphics[width=0.47\textwidth]{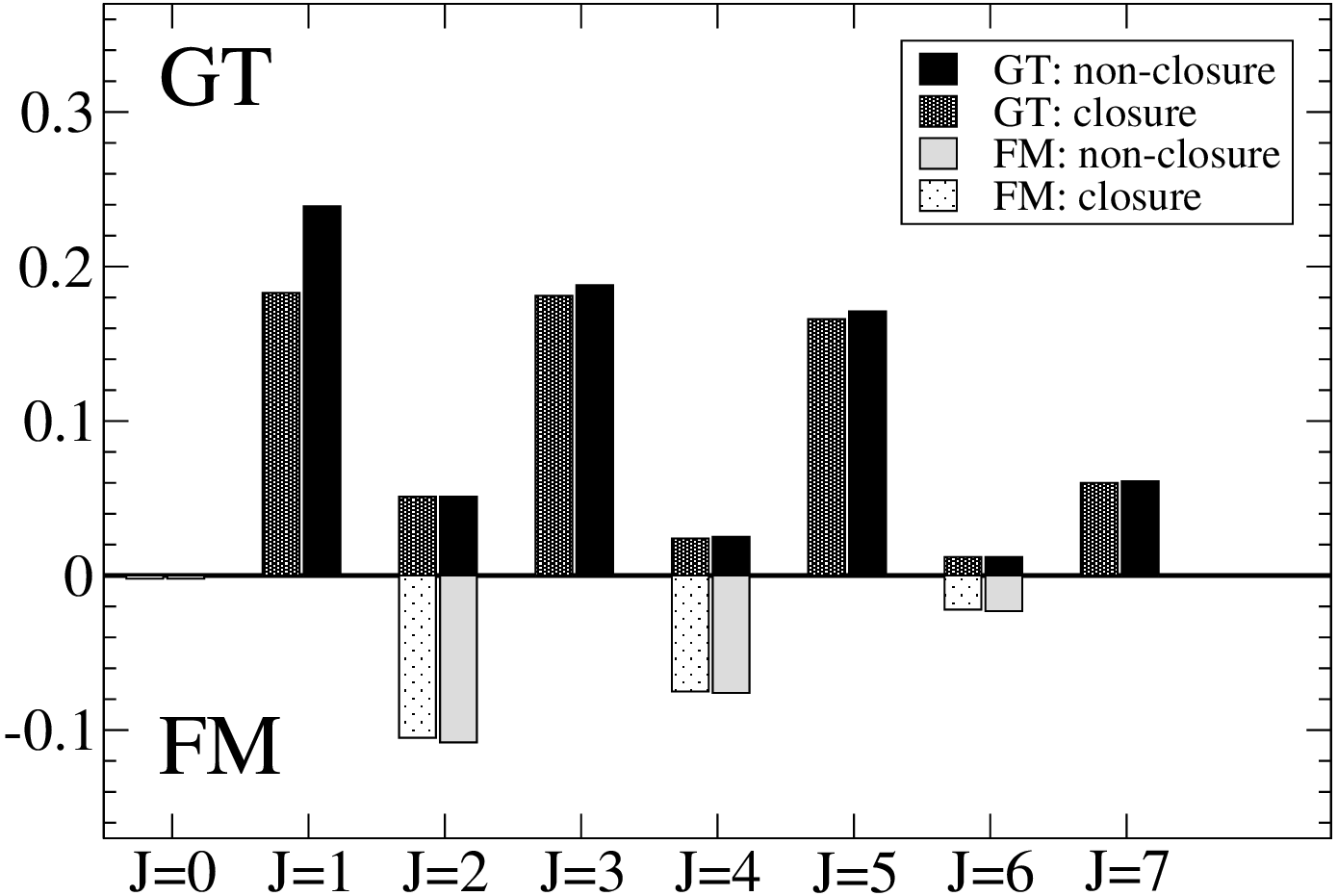}
\caption{Nonclosure vs. closure GT and F nuclear matrix elements 
for $0\nu\beta\beta$ decay of ${}^{48}$Ca for different 
spins $J$ of the intermediate states $|\kappa \rangle$. 
Solid black and gray bars correspond to the 
nonclosure approach, while shaded bars represent 
the closure approximation.\\
}\label{fig3}
\end{center}
\end{figure}

Figure \ref{fig3} and Table \ref{tbl1} present the
comparison of the results for the nonclosure approach, Eq. (\ref{nme4}), with the closure NME, 
for the decay of ${}^{48}$Ca. 
In these calculations 
 we use
\beq \label{eq14}
\left[ E_\kappa - (E_i + E_f)/2 \right] \rightarrow 1.9\,{\rm MeV}
 + E^{*}_\kappa,
\eeq
where $E^{*}_\kappa$ is the excitation energy of the intermediate nucleus ${}^{48}$Sc, 
the harmonic oscillator parameter $b_{osc}=1.989\,{\rm fm}$, and for the closure approximation the average energy 
was $\langle E \rangle = 7.72\,{\rm MeV}$.
Here, we also used the AV18 SRC parametrization \cite{prc10}.
In Fig. \ref{fig3} the nonclosure NME are represented by
solid black and gray bars and the closure NME are the dashed bars, 
shown for various angular momenta $J_\kappa$ of intermediate 
states $|\kappa\rangle$. 
The Gamow-Teller matrix elements are all positive (upper part),
and the Fermi matrix elements are all negative (bottom part). 
The main difference between closure and nonclosure comes from the 
GT nuclear matrix element corresponding to the intermediate angular momentum $J_\kappa = 1$. 
The reason is that the transitions from an initial $0^+$ state to an intermediate $1^+$ state occur most naturally via 
the $\bsig \tau_-$ operator. 
For the other types of operators and for the intermediate spins 
different from $J_\kappa=1$,
we have to expand the form factors over the neutrino momentum $q$, 
which makes the nuclear matrix element insensitive to low excitation energies, and therefore does not contribute
to the difference between closure and nonclosure NME.
This decomposition of the matrix elements, which is often provided by QRPA
calculations (see, e.g., Fig. 3 of Ref. \cite{qrpa-J}) 
is presented for the first time here as a result of a shell-model analysis.
As mentioned in Ref. \cite{prc10}, there are no contributions 
from the negative-parity states of the intermediate
nucleus when the model space is restricted to one major harmonic oscillator shell.

\begin{figure}
\begin{center}
\includegraphics[width=0.45\textwidth]{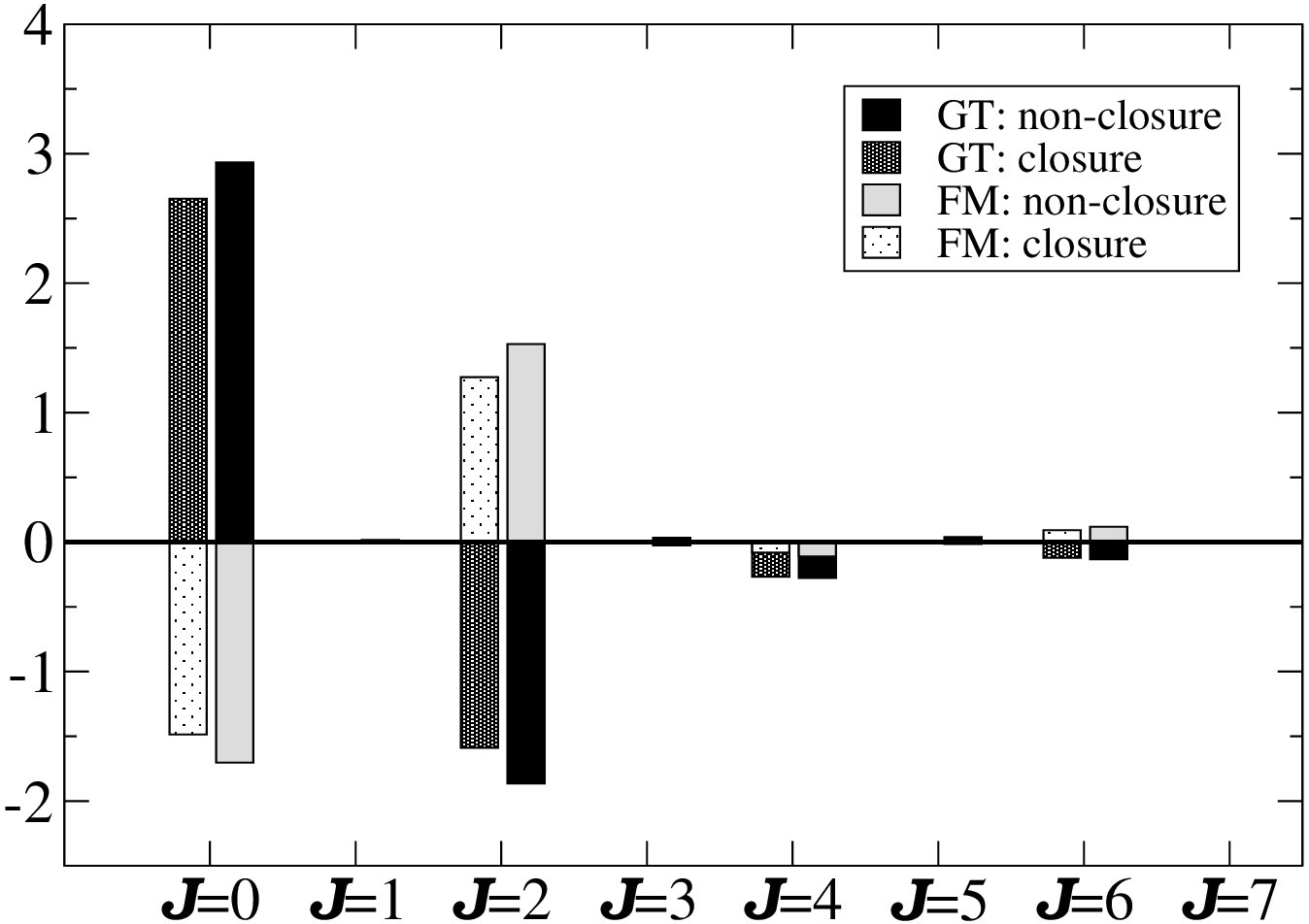}
\caption{Nonclosure vs. closure GT and F nuclear matrix elements 
for $0\nu\beta\beta$ decay of ${}^{48}$Ca calculated for 
certain spins of two initial neutrons and two final protons: 
$\langle 13, {\cal J}| {\cal O}^\alpha | 24, {\cal J}\rangle$. 
The coloring scheme and parameters are the same as in Fig. \ref{fig3}. \\}
\label{fig4}
\end{center}
\end{figure}

Figure \ref{fig4} represents another possible way to decouple the nuclear matrix elements. In this approach we consider two-body matrix elements
$\langle 1 3 | {\cal O}_\alpha | 2 4 \rangle$ where 
the single-particle states $|1\rangle$ and $|3\rangle$ (proton states) 
and the states $|2\rangle$ and $|4\rangle$ (neutron states) are coupled 
to certain common spin ${\cal J}$, so that the total NME can be presented as $M_\alpha = \sum_{\cal J} M_\alpha({\cal J})$. 
The details of such decoupling are in Appendix \ref{ap2}. 
The nonclosure NME in Fig. \ref{fig4} are represented with solid
 black and gray bars and the closure NME are the dashed bars. 
In contrast to the intermediate spin decoupling, where all the spins 
$J_\kappa$ contribute coherently (see Fig. \ref{fig3}),
in the ${\cal J}$-decoupling scheme we see a significant cancellation between ${\cal J}=0$ and ${\cal J}=2$.
Such a cancellation is responsible for the small matrix element of the double magic nucleus ${}^{48}$Ca. 
Similar effects have been observed
in seniority-truncation studies of the NME of 
${}^{48}$Ca \cite{menendez-sen} (see also Ref. \cite{npa818} for effects of higher seniority in shell model calculations). QRPA results are available for
heavier nuclei (see, e.g., Fig. 1 of Ref. \cite{qrpa-J}), for which the 
${\cal J}=0$ and ${\cal J}=2$ contributions are still dominant, but the cancellation effect is significantly reduced.

\begin{figure}
\begin{center}
\includegraphics[width=0.48\textwidth]{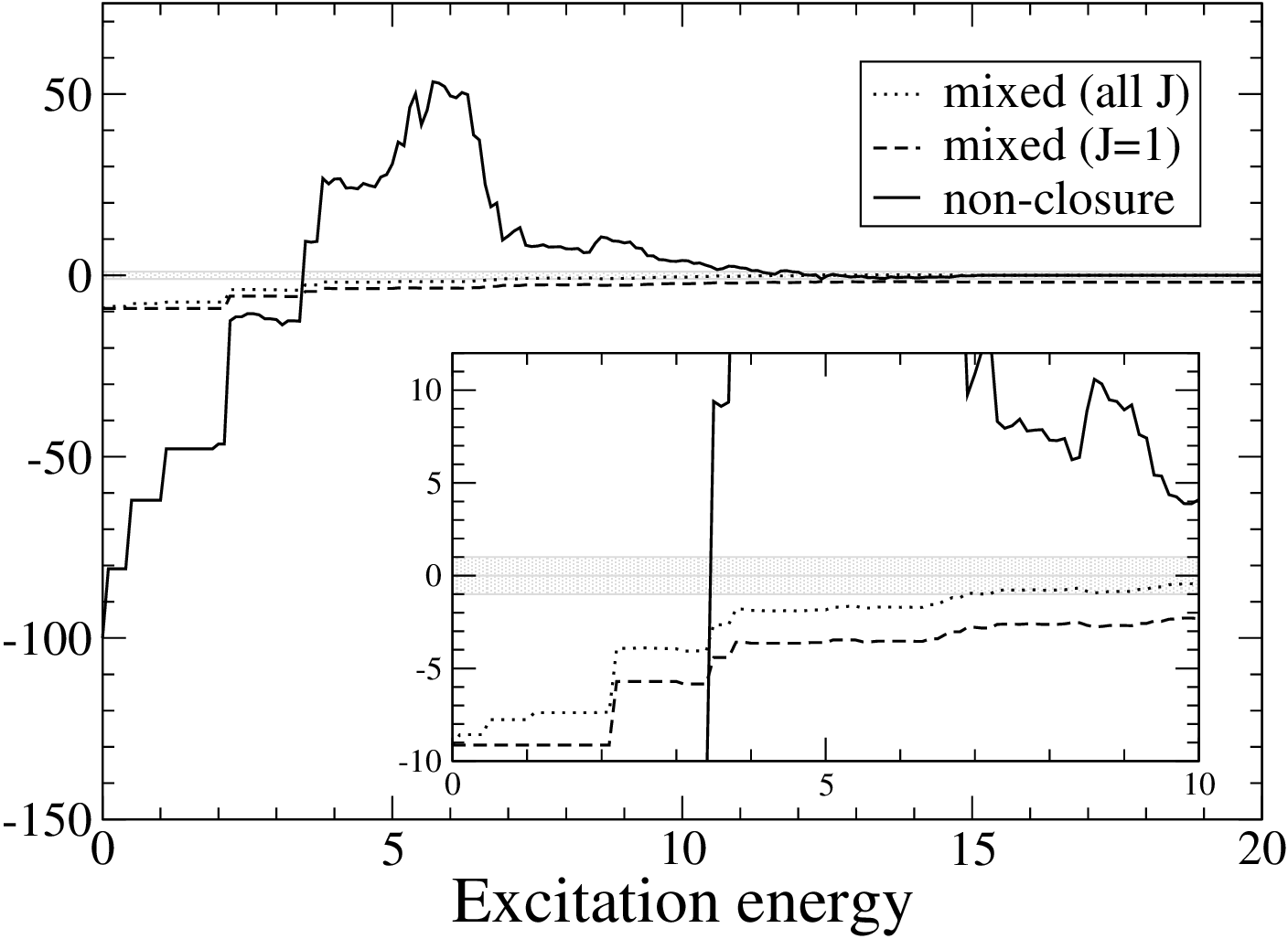}
\caption{
Convergence of total NME for $0\nu\beta\beta$ decay 
of ${}^{48}$Ca to its final value, $100\% \times \delta M/M$,
as a function of the cutoff energy. 
New mixed methods of calculation (presented by 
dotted and dashed lines) have much better convergence
compared to the pure nonclosure approach (solid line). 
The insert shows the low-energy part.\\
}\label{fig5}
\end{center}
\end{figure}

Figure \ref{fig5} presents the convergence of the total nuclear matrix element for ${}^{48}$Ca to its final value, $100\% \times \delta M/M$,
as a function of the cutoff energy.
The solid line defined by 
Eq. (\ref{nme3})  represents the nonclosure approach. We see that the matrix elements approach their final values 
(with the central shaded region corresponding to $\pm 1\%$) quite fast. 
In order to calculate the sum over the intermediate states in Eq. (\ref{nme2}) within an accuracy better than 1\% it is enough to include only the first 100 states for each $J_\kappa$. We conclude that if we restrict the sum over intermediate states to about 100 states of each spin, the uncertainty we introduce into the calculation by this restriction would be of the order of 1\%. 

The dotted and dashed lines in Fig. \ref{fig5} represent the mixed method, where the NME are defined by Eq. (\ref{nme5}). The dotted lines show the total matrix element, which includes all possible intermediate spins $J_\kappa$. It converges much faster than the pure nonclosure matrix element. 
To get an accuracy of about 1\% using this method we have to take into account only states of up to 7 MeV in excitation energy (about 20 states per each $J_\kappa$). 
The hope is that using this mixed method we can achieve the
desirable accuracy significantly faster (with a lower number of intermediate states) than using a pure  nonclosure approach. To obtain the NME of heavier nuclei, for which the dimensions are extremely high, such 
a decrease in computational demands can be crucially important. 

The main contribution to the NME originates from the intermediate states with
spin $J_\kappa=1$ (see Fig. \ref{fig3}). This observation can be used to decrease the number of intermediate states required for a given accuracy.
The dashed lines in Fig. \ref{fig5} represent the NME when the intermediate sates with $J_\kappa=1$ are only
taken into account. The difference between dotted and dashed lines is only 2\%, which means that if 
we include only the first 20 states with $J_\kappa=1$ we already achieve an accuracy of 3\%. This allows us to avoid calculation of all the intermediate states with $J_\kappa \ne 1$ and still get the NME with good accuracy.  

Table \ref{tbl1} summarizes the difference between the total
 matrix elements calculated within the closure approximation and the nonclosure
approach. We found about an 11\% percent difference for the GT matrix element, which is quite noticeable. For the total matrix element this difference
decreases to 10\%. 

The nonclosure results can be obtained in the closure approximation if
one uses an appropriate value for $\langle E \rangle$ and not 
$\langle E \rangle = 7.72$ MeV as suggested by QRPA calculations 
\cite{tomoda}. For CD-Bonn and AV18 SRC parametrizations 
(see Table \ref{tbl2}) this appropriate energy is found to be about
$\langle E \rangle = 0.5$ MeV, but its value may be different
for different model spaces, interactions, or SRC parametrizations.   

\begin{table}[ht]
\begin{center}
\begin{tabular}{||l|c|c|c||}
\hhline{|t:=:=:=:=:t|}
\hhline{||----||}
\rule{0cm}{0.33cm}   & Closure & nonclosure & $\delta M / M$\\
\hhline{||----||}
\rule{0cm}{0.33cm}  Gamow-Teller, $M^{0\nu}_{GT}$ & 0.676 & 0.747 & 11\% \\
\hhline{||----||}
\rule{0cm}{0.33cm}  Fermi, $M^{0\nu}_F$ &  -0.204 & -0.208 & 2\% \\
\hhline{||----||}
\rule{0cm}{0.33cm}  Tensor, $M^{0\nu}_T$ & -0.077 & -0.079 & 3\% \\
\hhline{||----||}
\rule{0cm}{0.33cm}  Total, $M^{0\nu}$ & 0.729 & 0.800 & 10\% \\ 
\hhline{|b:=:=:=:=:b|}
\end{tabular}
\caption{Nonclosure vs closure nuclear matrix elements 
for $0\nu\beta\beta$ decay of ${}^{48}$Ca
calculated for the AV18 SRC parametrization and with closure average 
energy $\langle E \rangle=7.72$ MeV.\\}\label{tbl1}
\end{center}
\end{table}

\begin{table}[ht]
\begin{center}
\begin{tabular}{||l|c|c|c|c||}
\hhline{|t:=:=:=:=:=:t|}
\hhline{||-----||}
\rule{0cm}{0.33cm} SRC  & $M^{0\nu}_{GT}$ & $M^{0\nu}_F$ & $M^{0\nu}_T$ & $M^{0\nu}$\\
\hhline{||-----||}
\rule{0cm}{0.33cm} None & 0.782 & -0.211 & -0.077 & 0.839\\
\hhline{||-----||}
\rule{0cm}{0.33cm} Miller-Spencer &  0.555 & -0.143 & -0.078 & 0.568\\
\hhline{||-----||}
\rule{0cm}{0.33cm} CD-Bonn & 0.810 & -0.226 & -0.079 & 0.875\\
\hhline{||-----||}
\rule{0cm}{0.33cm} AV18 & 0.747 & -0.208 & -0.079 & 0.800\\
\hhline{|b:=:=:=:=:=:b|}
\end{tabular}
\caption{Nonclosure nuclear matrix elements 
for $0\nu\beta\beta$ decay of ${}^{48}$Ca 
calculated for different SRC parametrizations 
\cite{prc10}. \\}\label{tbl2}
\end{center}
\end{table}

Finally, Table \ref{tbl2} presents the nonclosure $^{48}$Ca NME calculations
performed with different SRC parametrization sets \cite{prc10}. 

\section{Conclusions and Outlook}

In conclusion, we investigated the closure versus nonclosure approach of the 
$0 \nu \beta \beta$ NME for $^{48}$Ca using for the first time shell-model techniques in the realistic $pf$ shell valence space.
We found that the closure approximation always gives smaller NME,
$M^{0 \nu}_{\nu}$, by about 10\%. 
A similar comparison of closure versus nonclosure NME for heavy
nuclei, such as $^{76}$Ge, $^{96}$Zr, $^{100}$Mo, and $^{130}$Te, 
was done within the QRPA method in Ref. \cite{simvo11} (see, e.g., its Fig. 4), 
where the authors came to the same conclusion, namely, that the
nonclosure NME are about 10\% larger than the closure NME.

In addition, we were able to obtain for the first time a decomposition 
of the shell-model NME versus the total spin $J$ of the intermediate 
states, and we found that for the case of $^{48}$Ca the $J=1$ 
states provide the largest contribution. We have also found that 
most of the additional difference between closure and 
nonclosure comes from the transitions to the $1^+$ states in the 
intermediate nucleus.

By combining the nonclosure and closure approaches together
we propose a new method of calculating the $0\nu\beta\beta$ NME, 
which converges very quickly using only a very small number of 
states in the intermediate nucleus. This result suggests that one 
can apply this method to obtain the shell-model nonclosure NME 
for $0\nu\beta\beta$ decay of heavier nuclei, such as 
${}^{76}$Ge or ${}^{82}$Se. It would be also interesting to go
beyond the closure approximation for the NME corresponding to 
other mechanisms that may contribute to the $0\nu\beta\beta$ decay
rates \cite{ves12,prc13,prl13}. 

Finally, it is worth mentioning that the nonclosure approach does 
not constrain the states of the intermediate nucleus to be in the same model 
space used for the initial and the final state, as is the case for 
the closure approximation (see, e.g., Ref. \cite{prc10}).
For example, it was recently shown \cite{prl13} that the two-neutrino 
double-$\beta$ decay NME, which need to be calculated using a 
nonclosure approach, could change if the model space used for the 
intermediate $1^+$ states is enlarged. This effect could be 
considered in future studies. Here, we use for the nonclosure approach 
the same constraint as that imposed by the closure approximation.

\vspace*{0.5cm}
RAS is grateful to N. Auerbach and V. Zelevinsky for constructive 
discussions. Support from  the NUCLEI SciDAC Collaboration under
U.S. Department of Energy Grant No. DE-SC0008529 is acknowledged. 
MH also acknowledges U.S. NSF Grant No. PHY-1068217.

\appendix

\section{Form Factors} \label{ap1}
The form factors $h_{\alpha}(q^2)$ in the neutrino potentials\\
given by Eq. (\ref{pot}) have the following form
\begin{align} 
h_{F}(q^2) = & \frac{g^2_V(q^2)}{g^2_V},  \notag \\
h_{GT}(q^2) = & \frac{g^2_A(q^2)}{g^2_A} 
\left[1- \frac{2}{3} \frac{q^2}{q^2+m^2_\pi} + 
\frac{1}{3} \left(\frac{q^2}{q^2+m^2_\pi}\right)^2 \right]  \notag \\
& +\frac{2}{3} \frac{g^2_M(q^2)}{g^2_A} \frac{q^2}{4 m^2_p},  \notag \\
h_{T}(q^2) =& \frac{g^2_A(q^2)}{g^2_A} 
\left[\frac{2}{3} \frac{q^2}{q^2+m^2_\pi} - 
\frac{1}{3} \left(\frac{q^2}{q^2+m^2_\pi}\right)^2 \right]   \notag \\
&  +\frac{1}{3} \frac{g^2_M(q^2)}{g^2_A} \frac{q^2}{4 m^2_p}. \label{ap1.1}
\end{align}
Here $g_V=1$ and $g_A=1.254$ are the vector and axial constants and the 
form factors $g_{V, A, M}(q^2)$ are given by
\beq
\begin{aligned}
g_{V}(q^2)  = & \frac{g_V}{(1+q^2/\Lambda^2_V)^2},  \\
g_{M}(q^2) = & (\mu_p-\mu_n) g_V(q^2),\\
g_{A}(q^2) = & \frac{g_A}{(1+q^2/\Lambda^2_A)^2},  
\end{aligned}
\eeq
where the finite-size parameters $\Lambda_V=850$ MeV, $\Lambda_A=1086$ MeV, 
and the magnetic moments $(\mu_p-\mu_n)=4.7$.

\section{Nuclear Matrix Elements}\label{ap2}
The total matrix element of $0\nu\beta\beta$ 
decay, Eq. (\ref{nme2}), is given by the sum
over all the intermediate states $|\kappa\rangle$: 
\beq \label{ap2.1}
M^{0\nu}_\alpha = \sum_\kappa M^\alpha_\kappa.
\eeq
We can introduce two different partial matrix elements, 
one of them corresponding to the sum over all intermediate 
states with certain spin $J_\kappa$,
\beq \label{ap2.2}
M^{0\nu}_\alpha(J)=
\mathop{\sum M^\alpha_\kappa}_{\kappa \; (J_\kappa=J)} 
\mbox{  and  }\; M^{0\nu}_\alpha = \sum_{J} M^{0\nu}_\alpha(J),
\eeq 
and the other one corresponding to the sum over all intermediate
states when the single-particle orbitals $|1\rangle$, $|3\rangle$ 
and $|2\rangle$, $|4\rangle$ in two-body matrix elements 
$\langle 1 3 | {\cal O}_\alpha | 2 4 \rangle$
are coupled into total spin ${\cal J}$ as
\beq \label{ap2.3}
|1 3, {\cal J M} \rangle = \sum_{m_1 m_3} C^{\cal J M}_{j_1 m_1 \; j_3 m_3} 
|j_1 m_1 \rangle |j_3 m_3 \rangle,
\eeq
so that
\beq \label{ap2.4}
M^{0\nu}_\alpha({\cal J})=
\mathop{\sum M^\alpha_\kappa}_{\kappa \; ({\cal J}={\rm fixed})} \mbox{  and  }
\; M^{0\nu}_\alpha = \sum_{{\cal J}} M^{0\nu}_\alpha({\cal J}). 
\eeq 
The nuclear matrix elements, $M^\alpha_\kappa$, which we need for 
Eqs. (\ref{ap2.2}) and (\ref{ap2.4}), can be obtained from 
\beq \label{ap2.5}
\begin{aligned}
M^\alpha_\kappa = & f_T \sum_{1234} 
\left[ (-1)^{j_2 +j_4+{\cal J}} \Pi_{J_\kappa J_\kappa {\cal J}} 
\rule{0cm}{0.4cm} \right. \\
& \times
\left\{
\begin{array}{ccc}
j_1 & j_2 & J_\kappa \\
j_4 & j_3 & {\cal J}
\end{array}
\right\} 
\langle 1 3, {\cal J} || {\cal O}_\alpha || 2 4, {\cal J} \rangle \\
& \left. \rule{0cm}{0.4cm} \times 
\rho_{21}(J_\kappa\; t, i\rightarrow\kappa) 
\rho_{34}(J_\kappa\; t, f\rightarrow\kappa)^* \right],
\end{aligned}
\eeq
where $\Pi_{a b \cdots z}=\sqrt{(2a+1)(2b+1)\cdots(2z+1)}$;
operators ${\cal O}_\alpha$ are defined by Eq. (\ref{op}) except for 
the isospin structure $\tau_{1-} \tau_{2-}$, which was taken into
account separately by the isospin factor $f_T$; and $\rho_{21}$ and $\rho_{34}$
are the one-body transitional densities (OBTD) to be defined below. 
Note that the two-body matrix elements in the above equation are 
unsymmetrized. 

\section{One-Body Transitional Densities} \label{ap3}
Nuclear initial, intermediate, and final states can be presented 
in the proton-neutron (PN) formalism or in the isospin (T) formalism. 

In the PN formalism the nuclear states have certain isospin projection 
but no certain isospin. The isospin factor in this case simply equals one:
\beq \label{ap3.1}
f_T=\langle p(1) p(3) | \tau_{1-} \tau_{2-} | n(2) n(4) \rangle = 1.
\eeq
For the OBTD we can ignore the isospin indices and get
\beq \label{ap3.2}
\rho_{2 1}(J, i \rightarrow \kappa) = \frac{1}{\sqrt{2 J + 1}}
\langle \kappa || [ \hat{c}^\dagger_{1} \otimes \tilde{\hat{c}}_2 ]_{J} || i \rangle,
\eeq
where the tilde denotes a time-conjugated state, $\tilde{\hat{c}}_{j m} = (-1)^{j+m}{\hat{c}}_{j -m}$. 

In the T formalism, the nuclear states have certain isospin, which
results in a non-trivial isospin factor, 
\beq \label{ap3.3}
f_T = - \frac{3}{2 T_\kappa +1}
{C^{T_\kappa {T_\kappa}_z}_{T_f {T_f}_z\, 1 +1}}
{C^{T_\kappa {T_\kappa}_z}_{T_i {T_i}_z\, 1 -1}},
\eeq
and a different definition of the OBTD,
\beq \label{ap3.4}
\rho_{2 1} (J t, i\rightarrow\kappa) = \frac{\langle \kappa ||| 
[ \hat{c}^\dagger_{1} \otimes \tilde{\hat{c}}_2 ]_{Jt} 
||| i \rangle}{\sqrt{2t+1}\sqrt{2J+1}},
\eeq
where $\langle ||| \cdots ||| \rangle$ stands for the reduced matrix 
element in both spin and isospin spaces, and the
time-conjugated state includes the additional factor
$\tilde{\hat{c}}_{\frac{1}{2} \tau}=(-1)^{\frac{1}{2}+\tau}\hat{c}_{\frac{1}{2} -\tau}$. 

\section{Reduced Matrix Elements} \label{ap4}

To calculate the reduced matrix elements in Eq. (\ref{ap2.5}), 
$ \langle 1 3, {\cal J} || {\cal O}_\alpha || 2 4, {\cal J} \rangle $,
we transform to relative and center-of-mass
coordinates ${\bf r} = {\bf r_1} - {\bf r_2}$ and 
${\bf R} = ({\bf r_1}+{\bf r_2})/2$. The operators ${\cal O}_\alpha$ 
depend only on relative coordinates, so let us rewrite
these operators in such a
form that will allow us to focus on the spin and coordinate dependencies 
(and for simplicity we omit here the isospin factor $\tau_{1-} \tau_{2-}$)
\beq \label{ap4.1}
{\cal O}_\alpha =  
\sum^c_{\gamma=-c} (-1)^{\gamma} \Sigma^{\alpha}_{c\,\; -\gamma}
\left\langle { {\cal A}^\alpha_{c\; \gamma}(q, {\bf r}) } 
\right\rangle_\kappa,
\eeq
where $c=0$ for $\alpha=\{GT,F\}$ and $c=2$ for $\alpha=T$. Here 
$\Sigma^\alpha_{c\; \gamma}$ include all the spin dependence as 
\beq \label{ap4.2}
\Sigma^{GT}_{00}=(\bsig_1 \cdot \bsig_2),\;\; \Sigma^{F}_{00}=1,\;\; 
\Sigma^{T}_{2 \gamma} = 
\left[ \sigma_1 \otimes \sigma_2 \right]_{2 \gamma}, 
\eeq
${\cal A}^\alpha_{c\; \gamma}$ carry the coordinate and $q$ dependence as
\beq \label{ap4.3}
\begin{aligned}
{\cal A}^{GT}_{0\; 0}(q, {\bf r}) = 
{\cal A}^{F}_{0\; 0}(q, {\bf r}) = j_0(qr), \\
{\cal A}^{T}_{2\; \gamma}(q, {\bf r}) = 
\sqrt{\frac{24\pi}{5}} j_2(qr) Y_{2 \gamma}({\bf n}),
\end{aligned}
\eeq
and the average over neutrino momentum $q$ means
\beq \label{ap4.4}
\left\langle \rule{0cm}{0.33cm} {\cal T}^\alpha(q) \right\rangle_\kappa = 
\frac{2 R}{\pi} \int {\cal T}^\alpha(q)
\frac{ h_\alpha(q^2) \, q d q}
{q + E_\kappa - (E_i + E_f)/2 },
\eeq
where ${\cal T}^\alpha(q)$ is an arbitrary function of $q$ that has a
certain index $\alpha=\{GT,F,T\}$, so that each function 
${\cal T}^\alpha(q)$ is averaged with its own form factor $h_\alpha(q^2)$. 
Now, omitting the average over the neutrino momentum, we can present 
the reduced matrix elements as
\beq \label{ap4.5}
\begin{aligned}
\langle 1 3, {\cal J} || &
\sum^c_{\gamma=-c} (-1)^{\gamma} \Sigma^{\alpha}_{c\,\; -\gamma} 
{\cal A}^\alpha_{c\; \gamma}(q, {\bf r})
|| 2 4, {\cal J} \rangle \\
 & = \Pi_{\cal J} \sum C_{13} 
 C_{24} (-1)^{S+\lambda' + {\cal J}} \\
 & \times \left\{
\begin{array}{ccc}
S' & S & c \\
\lambda & \lambda' & {\cal J}
\end{array}
\right\} \langle S || \Sigma^{\alpha}_c || S' \rangle
\langle \lambda || {\cal A}^\alpha_{c} || \lambda' \rangle,
\end{aligned} 
\eeq
where the coefficients $C_{13}$ and $C_{24}$ are responsible for
coupling the nucleon individual spins and angular momenta 
to certain common spin and angular momentum:
\beq \label{ap4.6}
\begin{aligned}
C_{13} = \langle S \lambda; {\cal J} |
 l_1 j_1, l_3 j_3; {\cal J} \rangle, \\
C_{24} = \langle S' \lambda'; {\cal J} | 
l_2 j_2, l_4 j_4; {\cal J} \rangle.
\end{aligned}
\eeq
They can be easily calculated from
\beq \label{ap4.7}
\begin{aligned}
C_{13} = \Pi_{j_1 j_3 \lambda S} \left\{ 
\begin{array}{ccc}
\frac{1}{2} & l_3 & j_3 \\
\frac{1}{2} & l_1 & j_1 \\
S & \lambda & {\cal J}
\end{array}
\right\}, \\
C_{24} = \Pi_{j_2 j_4 \lambda' S'} \left\{ 
\begin{array}{ccc}
\frac{1}{2} & l_4 & j_4 \\
\frac{1}{2} & l_2 & j_2 \\
S' & \lambda' & {\cal J}
\end{array}
\right\}.
\end{aligned}
\eeq  
Calculation of the spin reduced matrix element in Eq. 
(\ref{ap4.5}) is straightforward, but the radial and angular parts 
require more attention. To transform to relative 
coordinate we need to use Talmi-Moshinsky brackets $D_{13}$ 
and $D_{24}$ 
\beq \label{ap4.8}
\begin{aligned}
\langle 13, \lambda || {\cal A}^{\alpha}_c || 24, \lambda' \rangle
= \Pi_{\lambda \lambda'} \sum D_{13} D_{24} (-1)^{L+\lambda'+l_r} \\
\times \left\{
\begin{array}{ccc}
l_r' & l_r & c\\
\lambda & \lambda' & L
\end{array}
\right\} \langle n_r l_r || {\cal A}^{\alpha}_c || n_r' l_r' \rangle,
\end{aligned} 
\eeq
where the sum runs over all allowed center-of-mass and relative radial and
angular quantum numbers: $\{N,L\}$, $\{n_r,l_r\}$, and $\{n_r',l_r'\}$.
Coefficients $D_{13}$ and $D_{24}$ perform transformation of the
orbital wave functions to the relative and center-of-mass wave functions
\beq \label{ap4.9}
\begin{aligned}
|n_1 l_1, n_3 l_3; \lambda \rangle = \sum_{n_r l_r, N L} 
D_{13} | n_r l_r, N L; \lambda \rangle, \\
|n_2 l_2, n_4 l_4; \lambda' \rangle = \sum_{n_r' l_r', N L} 
D_{24} | n_r' l_r', N L; \lambda' \rangle.
\end{aligned}
\eeq 
The angular reduced matrix elements in 
Eq. (\ref{ap4.8}) have a standard form and can be 
found with the help of Ref. \cite{varsh}, and the radial part of the reduced 
matrix elements can be integrated analytically, which allows us to 
significantly increase the accuracy and efficiency of the calculations. 
Indeed, the radial matrix elements we are 
interested in Eq. (\ref{ap4.8}) are
\beq \label{ap4.10}
\begin{aligned}
\langle n_r l_r | j_l(q r) | n_r' l_r' \rangle \\
=\int_{0}^{\infty} R_{n_r l_r} (r) j_l(q r) R_{n_r' l_r'} (r) r^2 d r,
\end{aligned}
\eeq
with $l=0,2$. They can be reduced to a sum of table integrals 
(see for example \cite{ryzhik}, p. 730, Eq. (6.631))
\beqn \label{ap4.11}
{\nu}^{\frac{m+1}{2}}
\nonumber
 \int_{0}^{\infty} r^m e^{- \nu r^2} j_l(q r) d r   \\
 = \frac{\sqrt{\pi}}{4} k! z^{{l}/{2}} L^{(l+\frac{1}{2})}_k(z) e^{-z},
\eeqn
where $k=(m-l-2)/2 $ (and in our case $k$ is always an integer and positive),
$z=q^2/4\nu$, and $L^{(l+\frac{1}{2})}_k(z)$ are generalized Laguerre polynomials.
To use these integrals one needs to  
expand the radial wave function, $R_{n l}$, in Eq. (\ref{ap4.10}). 
We used the standard expansion of generalized Laguerre polynomials
\beq \label{ap4.12}
 L^{(\beta)}_n(\nu r^2) = \sum_{i=0}^n \left( 
\begin{array}{c}
\beta + n \\
n - i
\end{array}
 \right)
 \frac{(-\nu r^2)^i}{i!}.
\eeq
The short range correlations are included by introducing the 
correlation function $f(r)$ that modifies the relative radial 
wave function at short distances (see, for example, Ref. \cite{prc10}),
\beq \label{ap4.13}
R_{n_r l_r}(r) \rightarrow [1+f(r)] R_{n_r l_r}(r).
\eeq
The function $ f(r) = -c e^{-ar^2} (1-br^2)$ 
is parametrized in such a way that we can still
integrate analytically the radial matrix elements 
with the help of relation (\ref{ap4.11}) (see \cite{prc86} and references 
therein).

Finally, the integration over the neutrino momentum $q$ was performed 
numerically by using Gauss-Laguerre and Gauss-Legendre quadrature rules.


\begin{thebibliography} {99}
\bibitem{corr1} The  small tensor term in Ref. \cite{prc13} has the opposite sign.
\end{thebibliography}
\end{document}